\documentclass[prb,aps,showpacs,onecolumn]{revtex4}
\usepackage{amssymb}
\usepackage{amsfonts}
\usepackage{amsmath}
\usepackage{graphicx}
\usepackage{bm}
\newcommand{\dprod}{\displaystyle\prod}

\begin{document}

\title{A short proof of the Gaillard-Matveev theorem based on shape
invariance arguments}
\author{Y.\ Grandati }
\affiliation{Equipe BioPhyStat, LCP A2MC, ICPMB, IF CNRS 2843, Universit\'{e} de
Lorraine-Site de Metz, 1 Bd Arago, 57078 Metz, Cedex 3, France}

\begin{abstract}
We propose a simple alternative proof of the Wronskian representation
formula obtained by Gaillard and Matveev for the\ Darboux-P\"{o}schl-Teller
(DPT) potentials. It rests on the use of singular Darboux-B\"{a}cklund
transformations applied to the free particle system combined to the shape
invariance properties of the DPT.
\end{abstract}

\maketitle

\section{\protect\bigskip Introduction}

In 2002, Gaillard and Matveev established an interesting connection between
the Darboux-Crum dressing formula and the Darboux-P\"{o}schl-Teller
potentials. More precisely, they proved the following theorem:

\bigskip \emph{Gaillard-Matveev's theorem:} The trigonometric Darboux-P\"{o}%
schl-Teller potential (TDPT) with integer parameters $m\geq n$ defined on $%
x\in \left] 0,\pi /2\right[ $ by

\begin{equation}
V(x;m,n)=\frac{m(m+1)}{\sin ^{2}x}+\frac{n(n+1)}{\cos ^{2}x},  \label{TDPT}
\end{equation}%
admits the following Wronskian representation

\begin{equation}
V(x;m,n)=-2\left( \ln W\left( u_{1},...,u_{m}\mid x\right) \right) ^{\prime
\prime },
\end{equation}%
where $W\left( u_{1},...,u_{m}\mid x\right) $ denotes the Wronskian of the
functions $u_{1}\left( x\right) ,...,u_{m}\left( x\right) $ which are given
by

\begin{equation}
\left\{ 
\begin{array}{c}
u_{k}\left( x\right) =\sin (lx),\text{ if }1\leq l\leq m-n \\ 
u_{m-n+l}\left( x\right) =\sin (\left( m-n+2l\right) x),\text{ if }1\leq
l\leq n.%
\end{array}%
\right.
\end{equation}

Gaillard and Matveev underlined that this result traduces the existence of a
chain of Darboux transformations connecting the constant potential to the
TDPT $V(x;m,n)$. If such a construction was already envisaged in Darboux's
seminal works \cite{darboux}, the theorem above precises what is the correct
choice of \textquotedblleft seeds\textquotedblright\ eigenfunctions of the
free Hamiltonian to reach this last potential.

The proof given in \cite{gaillard1} and \cite{gaillard2} rests on direct
evaluations of the Wronskian $W\left( u_{1},...,u_{m}\mid x\right) $ and is
quite sophisticated and technically involved. In this letter, we show that
this result can be obtained in a shorter way directly by following Darboux's
original idea. In a modern formulation the proof makes use of shape
invariance and SUSY QM arguments associated to singular Darboux-B\"{a}cklund
Transformations (DBT) built from excited eigenstates. If the use of singular
DBT has already been envisaged in several papers \cite%
{casahorran,panigrahi,robnik,marquez}, they are generally considered of less
interest, although recently proven to be crucial in the study of some
two-dimensional superintegrable systems \cite{marquette}.

We also examine the case of the Bessel potentials, which is the first
example envisaged by Darboux \cite{darboux}. This is a confluent case which
necessitate to employ Matveev's generalized Wronskian leading then to a
Wronskian version of the Rayleigh formula for Bessel functions.

\section{\protect\bigskip Darboux-B\"{a}cklund Transformations}

We begin to briefly recall ther essential features of DBT and shape
invariance. If $\psi _{\lambda }(x;a)$ is an eigenstate of the hamiltonian $%
\widehat{H}(a)=-d^{2}/dx^{2}+V(x;a),\ x\in I\subset \mathbb{R},$ ($a\in 
\mathbb{R}^{m}$ being a multiparameter) associated to the eigenvalue $%
E_{\lambda }(a)$ ($E_{0}(a)=0$)

\begin{equation}
\psi _{\lambda }^{\prime \prime }(x;a)+\left( E_{\lambda }(a)-V(x;a)\right)
\psi _{\lambda }(x;a)=0,  \label{EdS}
\end{equation}
then the Riccati-Schr\"{o}dinger (RS) function $w_{\lambda }(x;a)=-\psi
_{\lambda }^{\prime }(x;a)/\psi _{\lambda }(x;a)$ satisfies the
corresponding Riccati-Schr\"{o}dinger (RS) equation \cite{grandati}

\begin{equation}
-w_{\lambda }^{\prime }(x;a)+w_{\lambda }^{2}(x;a)=V(x;a)-E_{\lambda }(a).
\label{edr4}
\end{equation}

The class of RS equations is preserved by a specific subset of the group $%
\mathcal{G}$ of smooth $SL(2,\mathbb{R})$-valued curves $Map(\mathbb{R},SL(2,%
\mathbb{R}))$ . These transformations, called Darboux-B\"{a}cklund
Transformations (DBT), are built from any solution $w_{\nu }(x;a)$ of the
initial RS equation Eq(\ref{edr4}) as \cite{grandati,Ramos,carinena2}

\begin{equation}
w_{\lambda }(x;a)\overset{A\left( w_{\nu }\right) }{\rightarrow }w_{\lambda
}^{\left( \nu \right) }(x;a)=-w_{\nu }(x;a)+\frac{E_{\lambda }(a)-E_{\nu }(a)%
}{w_{\nu }(x;a)-w_{\lambda }(x;a)},  \label{transfoback2}
\end{equation}%
where $E_{\lambda }(a)\neq E_{\nu }(a)$. $w_{\lambda }^{\left( \nu \right) }$
is then a solution of the RS equation:

\begin{equation}
-w_{\lambda }^{\left( \nu \right) \prime }(x;a)+\left( w_{\lambda }^{(\nu
)}(x;a)\right) ^{2}=V^{\left( \nu \right) }(x;a)-E_{\lambda }(a),
\label{eqtransform}
\end{equation}%
with the same energy $E_{\lambda }(a)$ as in Eq(\ref{edr4}) but with a
modified potential

\begin{equation}
V^{\left( \nu \right) }(x;a)=V(x;a)+2w_{\nu }^{\prime }(x;a)
\label{pottrans}
\end{equation}%
called an extension of $V$. If $V^{\left( \nu \right) }$ is a rational
function of an a priori given variable ($x,$ $\tan x,$ $\sin x,$ $\cosh x...$%
) we call it a rational extension of $V$.

This can be schematically summarized as

\begin{equation}
\left\{ 
\begin{array}{c}
w_{\lambda }\overset{A(w_{\nu })}{\rightarrowtail }w_{\lambda }^{\left( \nu
\right) } \\ 
V\overset{A(w_{\nu })}{\rightarrowtail }V^{\left( \nu \right) }.%
\end{array}%
\right.  \label{schema}
\end{equation}

The corresponding eigenstate of $\widehat{H}^{\left( \nu \right)
}(a)=-d^{2}/dx^{2}+V^{\left( \nu \right) }(x;a)$ can be written

\begin{equation}
\psi _{\lambda }^{\left( \nu \right) }(x;a)=\exp \left( -\int dxw_{\lambda
}^{(\nu )}(x;a)\right) \sim \frac{1}{\sqrt{E_{\lambda }\left( a\right)
-E_{\nu }(a)}}\widehat{A}\left( w_{\nu }\right) \psi _{\lambda }(x;a),
\label{foDBT}
\end{equation}%
where $\widehat{A}\left( w_{\nu }\right) $ is a first order operator given
by $\widehat{A}\left( w_{\nu }\right) =d/dx+w_{\nu }(x;a).$ Eq(\ref{foDBT})
can still be written as

\begin{equation}
\psi _{\lambda }^{\left( \nu \right) }(x;a)\sim \frac{W\left( \psi _{\nu
},\psi _{\lambda }\mid x\right) }{\psi _{\nu }(x;a)},  \label{foDBTwronsk}
\end{equation}%
where $W\left( y_{1},...,y_{n}\mid x\right) $ is the Wronskian of the
functions $y_{1},...,y_{n}$.

From $V$, the DBT generates a new potential $V^{\left( \nu \right) }$
(quasi) isospectral to the original one and its eigenfunctions are directly
obtained from those of $V$ via Eq(\ref{foDBT}). Nevertheless, in general, $%
w_{\nu }(x;a)$ and then the transformed potential $V^{\left( \nu \right)
}(x;a)$ is singular at the nodes of $\psi _{\nu }(x;a)$. For instance, if $%
\widehat{H}(a)$ admits a bound state spectrum $\left( E_{n},\psi _{n}\right)
_{n\in \mathbb{N}}$ $(x;a)$, $V^{\left( n\right) }$ is regular only when $n=0
$, that is when $\psi _{n=0}$ is the ground state of $\widehat{H}$, and we
recover the usual SUSY partnership in quantum mechanics \cite{cooper,Dutt}.
Note that $A(w_{0})$ is a "state-deleting" DBT, the spectrum of the
superpartner hamiltonian $\widehat{\widetilde{H}}=\widehat{H}^{\left(
0\right) }$ having for fundamental level and ground state $E_{1}$ and $\psi
_{1}^{\left( 0\right) }$ respectively.

$V$ is said to be a translationally shape invariant potential (TSIP) if its
superpartner has the form

\begin{equation}
\widetilde{V}(x;a)=V^{\left( 0\right) }(x;a)=V(x;a+\alpha )+R(a),
\end{equation}%
where $\alpha \in \mathbb{R}^{m}$.

The question of successive iterations of DBT is very natural and is at the
center of the construction of the hierarchy of hamiltonians in the usual
SUSY QM scheme \cite{sukumar2}. Staying at the formal level, it can be
simply described by the following generalization of Eq(\ref{schema}) ($N_{j}$
denotes the j-uple $\left( \nu _{1},,...,\nu _{j}\right) $, with $N_{1}=\nu
_{1}$ and $N_{m}=\left( N_{m-1},\nu _{m}\right) $)

\begin{equation}
\left\{ 
\begin{array}{c}
w_{\lambda }\overset{A(w_{\nu _{1}})}{\rightarrowtail }w_{\lambda }^{\left(
\nu _{1}\right) }\overset{A(w_{\nu _{2}}^{\left( N_{1}\right) })}{%
\rightarrowtail }w_{\lambda }^{\left( N_{2}\right) }...\overset{A(w_{\nu
_{m}}^{\left( N_{m-1}\right) })}{\rightarrowtail }w_{\lambda }^{\left(
N_{m}\right) } \\ 
V\overset{A(w_{\nu _{1}})}{\rightarrowtail }V^{\left( \nu _{1}\right) }%
\overset{A(w_{\nu _{2}}^{\left( N_{1}\right) })}{\rightarrowtail }V^{\left(
N_{2}\right) }...\overset{A(w_{\nu _{m}}^{\left( N_{m-1}\right) })}{%
\rightarrowtail }V^{\left( N_{m}\right) },%
\end{array}%
\right.  \label{diagn}
\end{equation}%
where $w_{\lambda }^{\left( N_{m}\right) }$ is a RS function associated to
the eigenvalue $E_{\lambda }$ of the potential

\begin{equation}
V^{\left( N_{m}\right) }(x;a)=V(x;a)+2\sum_{j=1}^{m}\left( w_{\nu
_{j}}^{\left( N_{j-1}\right) }(x;a)\right) ^{\prime }.  \label{potnstep}
\end{equation}

The corresponding eigenfunction is given by (cf Eq(\ref{foDBT}) and Eq(\ref%
{foDBTwronsk}))

\begin{equation}
\psi _{\lambda }^{\left( N_{m}\right) }(x;a)=\widehat{A}\left( w_{\nu
_{m}}^{\left( N_{m-1}\right) }\right) \psi _{\lambda }^{\left(
N_{m-1}\right) }(x;a)=\frac{W\left( \psi _{\nu _{m}}^{\left( N_{m-1}\right)
},\psi _{\lambda }^{\left( N_{m-1}\right) }\mid x\right) }{\psi _{\nu
_{m}}^{\left( N_{m-1}\right) }(x;a)}.  \label{etats n}
\end{equation}%
or

\begin{equation}
\psi _{\lambda }^{\left( N_{m}\right) }(x;a)=\widehat{A}\left( w_{\nu
_{m}}^{\left( N_{m-1}\right) }\right) ...\widehat{A}\left( w_{\nu
_{1}}\right) \psi _{\lambda }(x;a).  \label{etats n2}
\end{equation}

By induction, we can prove the Crum formulas \cite{crum,Matveev}

\begin{equation}
\left\{ 
\begin{array}{c}
\psi _{\lambda }^{\left( N_{m}\right) }(x;a)=W\left( \psi _{\nu
_{1}},...,\psi _{\nu _{m}},\psi _{\lambda }\mid x\right) /W\left( \psi _{\nu
_{1}},...,\psi _{\nu _{m}}\mid x\right)  \\ 
V^{\left( N_{m}\right) }(x;a)=V(x;a)-2\left( \log W\left( \psi _{\nu
_{1}},...,\psi _{\nu _{m}}\mid x\right) \right) ^{\prime \prime }.%
\end{array}%
\right.   \label{crum}
\end{equation}

A succession of $m$ SUSY QM partnerships gives rise to a hierarchy of
regular potentials $V^{\left( N_{j}\right) },$ $j=1,...,m$, associated to
the m-uple $N_{m}=\left( 0,,...,m-1\right) $ , that is, to a set of seeds
functions $\left( \psi _{0},...,\psi _{m-1}\right) $ constituted by the $m$\
first bound states of the initial hamiltonian. In the case where the initial
potential is a TSIP, we obtain simply

\begin{equation}
V^{\left( N_{m}\right) }(x;a)=V(x;a)-2\left( \log W\left( \psi _{0},...,\psi
_{m-1}\mid x\right) \right) ^{\prime \prime }=V(x;a+m\alpha
)+\sum_{j=0}^{m-1}R(a_{j}),  \label{chainTSIP}
\end{equation}%
with

\begin{equation}
\psi _{k+m}^{\left( N_{m}\right) }(x;a)=\frac{W\left( \psi _{0},...,\psi
_{m-1},\psi _{k+m}\mid x\right) }{W\left( \psi _{0},...,\psi _{m-1}\mid
x\right) },\quad k\geq 0,  \label{chainTSIP2}
\end{equation}%
which is the $k$-th excited state of $V^{\left( N_{m}\right) }$ with the
associated energy $E_{k+m}$.

\section{From the constant potential to the TDPT potential}

Following Darboux \cite{darboux}, our starting point is the constant
potential on the real line $V(x)=0,$ $x\in \mathbb{R}.$ The physical energy
spectrum of the corresponding hamiltonian $\widehat{H}$ is constituted by an
open band $\mathcal{E}_{k}=k^{2}\in \left] 0,+\infty \right[ ,\ k>0$, where
the two-dimensional eigenspace  associated to $\mathcal{E}_{k}$ is spanned
by the (unnormalized) scattering eigenstates

\begin{equation}
\phi _{k}^{odd}\left( x\right) =\sin (kx)/k,\ \phi _{k}^{even}\left(
x\right) =\cos (kx)/k.  \label{eppotconst}
\end{equation}

We then build a chain of extensions of $V$ where the first DBT we choose is
based on a scattering state satisfying the Dirichlet boundary condition at
the origin, namely

\begin{equation}
\psi _{0}\left( x\right) =\phi _{1}^{odd}\left( x\right) =\sin \left(
x\right) ,\ w_{0}\left( x\right) =-\cot \left( x\right) ,
\end{equation}%
where the choice of the value $k=1$ is for pure convenience. The first
extended potential is then

\begin{equation}
V^{\left( 0\right) }(x)=V(x)+2w_{0}^{\prime }(x)=\frac{2}{\sin ^{2}x}
\label{V0}
\end{equation}%
and the corresponding Schr\"{o}dinger equation

\begin{equation}
\left( -d^{2}/dx^{2}+V^{\left( 0\right) }(x)\right) \psi (x)=E\psi (x).
\end{equation}

At this step, Darboux \cite{darboux} writes

\textit{\textquotedblleft Pour des valeurs particuli\`{e}res de }$E$\textit{%
, cette \'{e}quation admet les solutions }$\sin ^{2}x,$\textit{\ }$\sin
^{2}x\cos x$\textit{. En les employant et en poursuivant l'application de la
m\'{e}thode, on parviendra successivement \`{a} des \'{e}quations de la forme%
}

\begin{equation*}
\left( -d^{2}/dx^{2}+\frac{n(n+1)}{\cos ^{2}x}+\frac{m(m+1)}{\sin ^{2}x}%
\right) \psi (x)=E\psi (x).\text{\textit{\textquotedblright }}
\end{equation*}

As emphasized by Gaillard and Matveev, the argument needs to be completed
and specified in particular to determine in a systematic way what is the
correct choice of the successive seeds functions to reach this equation.

First note that Darboux's procedure does not consider the spectral
properties of the built hamiltonians which can then become more singular at
each step. This is the case of $V^{\left( 0\right) }$ which is singular at
every node $\xi _{l}=l\pi $ of $\psi _{0}\left( x\right) $. The
corresponding singularities are of the centrifugal barrier $\lambda /\left(
x-\xi _{l}\right) ^{2}$ with $\lambda >3/4$, that is, strong singularities
for which the transmission probability is zero \cite%
{frank,lathouwers,panigrahi}. This confines the particle in a single
interval $\left] \xi _{l},\xi _{l+1}\right[ $. If we take for instance $l=0$%
, the existence of the singularities in $0$ and $\pi $ can then be
considered as equivalent to the addition of Dirichlet boundary conditions at
the origin and at $x=\pi $. 
\begin{equation}
\psi \left( 0^{+}\right) =\psi \left( \pi ^{-}\right) =0.  \label{DBC}
\end{equation}

These lasts act as a filter which select among all the formal eigenfunctions 
$\psi _{k}^{\left( 0\right) }$ only those which satisfy Eq(\ref{DBC}). The
question is now to determine what are the appropriate values of $k$ which
allows to verify this constraint.

From a more general point of view, consider the problem to determine what is
the condition to which the solution $\psi _{\lambda }(x)$ of Eq(\ref{EdS})
is submitted in order that its image $\psi _{\lambda }^{\left( \nu \right) }$
via the DBT $A\left( w_{\nu }\right) $ satifies

\begin{equation}
\psi _{\lambda }(x_{0})=0,  \label{zero}
\end{equation}%
where $x_{0}\in I$ is a zero of $\psi _{\nu }$. In the vinicity of $x_{0}$
we can write

\begin{equation}
\psi _{\mu }(x)=a_{\mu ,0}+a_{\mu ,1}\left( x-x_{0}\right) +a_{\mu ,2}\left(
x-x_{0}\right) ^{2}+O\left( \left( x-x_{0}\right) ^{3}\right) .
\end{equation}

$x_{0}$ being a simple zero of $\psi _{\nu }$ we have $a_{\nu ,0}=0$ and $%
a_{\nu ,1}\neq 0$, which gives

\begin{equation}
w_{\nu }(x)\underset{x\rightarrow x_{0}}{\simeq }\frac{-1}{x-x_{0}}-\frac{%
a_{\nu ,2}}{a_{\nu ,1}}+O(\left( x-x_{0}\right) ).
\end{equation}

Inserting these results in Eq(\ref{foDBTwronsk}), we obtain%
\begin{equation}
\psi _{\lambda }^{\left( \nu \right) }(x)\underset{x\rightarrow x_{0}}{\sim }%
\frac{-a_{\lambda ,0}}{x-x_{0}}-\frac{a_{\nu ,2}}{a_{\nu ,1}}a_{\lambda
,0}+O(\left( x-x_{0}\right) )
\end{equation}%
and the condition (\ref{zero}) is satisfied iff $a_{\lambda ,0}=0$, that is,
iff $x_{0}$ is also a zero of $\psi _{\lambda }$.

In brief, a DBT $A\left( w_{\nu }\right) $ associated to a seed function $%
\psi _{\nu }$ having nodes on the definition interval generates an extended
hamiltonian which is singular on this interval. The restriction of this
extension between two singularities has a spectrum which consists in a
filtered version of the initial hamiltonian spectrum: we only keep the
levels $E_{\lambda }$ for which the initial eigenstates $\psi _{\lambda }$
have nodes at the singular points. This point of view is equivalent to the
one of Marquez et al \cite{marquez}.

\bigskip Applying this analysis to to the system considered above, we obtain
that the potential $V^{\left( 0\right) }$ restricted to the positive half
line and subjected to the boundary conditions Eq(\ref{DBC}) has a discrete
non degenerate spectrum given by $E_{l},$ $l\in \mathbb{N}^{\ast },$ with
the corresponding eigenstates

\begin{equation}
\psi _{l}^{\left( 0\right) }\left( x\right) =\widehat{A}\left( w_{0}\right)
\psi _{l}\left( x\right) \sim \left( l+1\right) \cos \left( \left(
l+1\right) x\right) -\cot x\sin \left( \left( l+1\right) x\right) ,
\label{ext1}
\end{equation}%
where we have noted $\psi _{l}\left( x\right) =\phi _{l+1}^{odd}\left(
x\right) $ and $E_{l}=\mathcal{E}_{l+1}$. In particular, the fundamental and
first excited eigenstates are $\psi _{1}^{\left( 0\right) }\left( x\right)
=\sin ^{2}\left( x\right) $ and $\psi _{2}^{\left( 0\right) }\left( x\right)
=\sin ^{2}x\cos x$. $V^{\left( 0\right) }$ can also be considered as the
image via $A\left( w_{0}\right) $ of the infinite square well

\begin{equation}
V_{SW}(x)=\left\{ 
\begin{array}{c}
0,\text{ if }x\in \left] 0,\pi \right[ \\ 
+\infty ,\text{ if }x\notin \left] 0,\pi \right[ .%
\end{array}%
\right.  \label{infwell}
\end{equation}

$V^{\left( 0\right) }$ being a particular TDPT potential (\ref{TDPT}) with $%
n=0,$ $m=1$, we can consider that $V_{SW}$ correspond to the limit case $%
V_{SW}(x)=V(x;0,0)=\underset{\mu ,\nu \rightarrow 0}{\lim }V(x;\mu ,\nu )$.

As it is well known \cite{cooper,Dutt,Gendenshtein,grandati}, the TDPT
potentials $V(x;\mu ,\nu )$ are second category translationally shape
invariant potentials (TSIP), the SUSY QM partner of $V(x;\mu ,\nu )$ being
given by

\begin{equation}
\widetilde{V}(x;\mu ,\nu )=V(x;\mu +1,\nu +1),  \label{SI1}
\end{equation}%
for $\mu ,\nu >0$ and

\begin{equation}
\left\{ 
\begin{array}{c}
\widetilde{V}(x;\mu ,0)=V(x;\mu +1,0),\quad \mu >0 \\ 
\widetilde{V}(x;0,\nu )=V(x;0,\nu +1),\quad \nu >0.%
\end{array}%
\right.  \label{SI2}
\end{equation}

Starting from $V^{\left( 0\right) }(x)=V(x;1,0)$, we can build a hierarchy
of $p-1$ extensions of via successive SUSY QM partnerships, that is, via a
chain of $p-1$ DBT based on the successive ground states. From Eq(\ref{SI2})
and Eq(\ref{chainTSIP}), we deduce that the final potential is $V(x;p,0)$
and can be seen as the end of a chain of $p$ extensions which starts from
the zero potential $V(x)$ and which is associated to the p-uple $%
N_{p}=(0,...,p-1)$, ie to the set of seeds eigenfunctions\bigskip\ $\left(
\psi _{0}\left( x\right) ,...,\psi _{p-1}\left( x\right) \right) =\left(
\sin \left( x\right) ,...,\sin \left( px\right) \right) $. Using a
straightforward recurrence, we deduce that the ground state of $V^{\left(
N_{p}\right) }(x)=V(x;p,0)=p(p+1)/\sin ^{2}x$ at energy $E_{p}$ is 
\begin{equation}
\psi _{p}^{\left( N_{p}\right) }\left( x\right) =\sin ^{p+1}x.  \label{fond}
\end{equation}

Since is even with respect to $\pi /2$, its eigenfunctions $\psi
_{n}^{\left( N_{p}\right) }$ are repectively symmetric (if $n-p$ is even) or
antisymmetric (if $n-p$ is odd) with respect to this point. An immediate
consequence is that its first excited eigenstate (at energy $E_{p+1}$) $\psi
_{p+1}^{\left( N_{p}\right) }$ has only one node on $\left] 0,\pi \right[ $
which is always located at $x=\pi /2$ independently of $p$. From Eq(\ref%
{fond}) we deduce

\begin{equation}
\psi _{p+1}^{\left( N_{p}\right) }\left( x\right) \sim \widehat{A}^{+}\left(
w_{p+1}^{\left( N_{p+1}\right) }\right) \psi _{p+1}^{\left( N_{p+1}\right)
}\left( x\right) \sim \cos x\sin ^{p+1}x.
\end{equation}

Suppose that we continue the chain of extension and build the following DBT
from this first excited eigenstate. The obtained potential is then singular
and presents a pole of second order at $x=\pi /2$. $\psi _{p+1}^{\left(
N_{p}\right) }$ being the image of $\psi _{p+1}(x)=\sin (\left( p+2\right)
x) $ by the chain associated to $N_{p}$

\begin{equation}
\psi _{p+1}^{\left( N_{p}\right) }(x)=\widehat{A}\left( w_{p}^{\left(
N_{p-1}\right) }\right) ...\widehat{A}\left( w_{0}\right) \psi _{p+1}(x),
\end{equation}%
the new chain is associated to the $\left( n+1\right) $-uple $\left(
N_{p},p+1\right) $ and the final extension is given by

\begin{equation}
V^{\left( N_{p},p+1\right) }(x)=\frac{2}{\cos ^{2}x}+\frac{\left( p+1\right)
(p+2)}{\sin ^{2}x}.
\end{equation}

The presence of the additional "strong" singularity in $\pi /2$ imposes a
Dirichlet boundary condition at this point. In the spectrum of the
restriction of $V^{\left( N_{p},p+1\right) }$ to $\left] 0,\pi /2\right[ $,
half of the levels $E_{n}^{\left( N_{p}\right) }$ of the preceding potential 
$V^{\left( N_{p}\right) }$ are eliminated. These are all those associated to
\textquotedblleft even\textquotedblright\ (with respect to $\pi /2$)
eigenstates, that is to the even values of the quantum number $n-p$.

The spectrum of $V^{\left( N_{p},p+1\right) }(x)$ contains then only the
levels $E_{p+1+2\left( j+1\right) },\ j\geq 0,$ and the associated
eigenstates $\psi _{p+1+2\left( j+1\right) }^{\left( N_{p},p+1\right) }$ are
the images of the initial eigenstates $\psi _{p+1+2\left( j+1\right) }$ via
the chain of BDT corresponding to the $\left( p+1\right) $-uple $\left(
N_{p},p+1\right) =(0,...,p-1,p+1)$. Starting from $V^{\left(
N_{p},p+1\right) }$ we can continue the chain by using standard SUSY QM
partnerships, all the extended potentials thus generated being perfectly
regular on $\left] 0,\pi /2\right[ $. But $V^{\left( N_{p},p+1\right) }$ is
also a TDPT

\begin{equation}
V^{\left( N_{p},p+1\right) }(x)=V(x;p+1,1).
\end{equation}

Consequently the shape invariance property Eq(\ref{SI1}) implies that after $%
q$ steps, we obtain as final extension

\begin{equation}
V^{\left( N_{p},p+1,...,p+q\right) }(x)=V(x;p+q,q).
\end{equation}

This shows in an explicite way what is exactly the chain of DBT which permit
to build the general TDPT potential of integer parameters as a rational (in $%
\sin x$) extension of the free particle system.

The Crum formula gives then

\begin{equation}
V(x;p+q,q)=-2\left( \ln W\left( \psi _{0},...,\psi _{p-1},\psi
_{p+1,}...,\psi _{p-1+2j},...,\psi _{p-1+2q}\mid x\right) \right) ^{\prime
\prime },
\end{equation}%
or, noting $l=p+q,$ 

\begin{eqnarray}
V(x;l,q) &=&\frac{q(q+1)}{\cos ^{2}x}+\frac{l(l+1)}{\sin ^{2}x}  \notag \\
&=&-2\left( \ln W\left( \sin x,...,\sin (\left( l-q\right) x),\sin \left(
(l-q+2)x\right) ,...,\sin \left( (l-q+2j)x\right) ,...,\sin \left( \left(
l+q\right) x\right) \mid x\right) \right) ^{\prime \prime },
\end{eqnarray}%
which is precisely the Gaillard-Matveev result.

\section{From the constant potential to the Bessel potential and a Wronskian
Rayleigh formula.}

To generate the TDPT potentials from the constant one, we start with a DBT
built from any (physical) diffusion state of the constant potential.
Nevertheless, we are not limited to this case and on a formal point of view
we can use any eigenfunction, physical or not to build this DBT. In
particular, we can choose as seed eigenfunction $\psi _{0}\left( x\right) =x=%
\underset{k\rightarrow 0}{\lim }\psi _{k}\left( x\right) $ whose associated
eigenvalue $E_{0}=0$ is located at the lower boundary of the physical
spectrum of $V$. The extended potential generated by the DBT $A(w_{0})$ is

\begin{equation}
V^{\left( 0\right) }(x)=2/x^{2}=V_{B}\left( x,1\right) ,
\end{equation}%
that is, the first Bessel potential of integer parameter, the general form
of which being $V_{B}\left( x;a\right) =a(a+1)/x^{2}.$

The two partner hamiltonians $\widehat{H}$ and $\widehat{H}^{\left( 0\right)
}$are strictly isospectral but the levels of are no more degenerated since
the strong singularity of $V^{\left( 0\right) }$ at the origin imposes to
retain in the spectrum of $\widehat{H}$ only the eigenstates which satisfy
the Dirichlet boundary condition at the origin 
\begin{equation}
\psi \left( 0^{+}\right) =0.  \label{DBC0}
\end{equation}

The physical spectrum of $\widehat{H}^{\left( 0\right) }$ is then

\begin{equation}
\left\{ 
\begin{array}{c}
E_{k}^{\left( 0\right) }=k^{2}\in \left] 0,+\infty \right[ ,\ k>0 \\ 
\psi _{k}^{\left( 0\right) }\left( x\right) =\frac{1}{E_{k}-E_{0}}\widehat{A}%
(w_{0})\psi _{k}\left( x\right) =\left( -\sin (kx)+kx\cos \left( kx\right)
\right) /k^{3}x,%
\end{array}%
\right.
\end{equation}%
the unphysical "fundamental" eigenfunction for $k=0$ being $\psi
_{0}^{\left( 0\right) }\left( x\right) =x^{2}\sim \underset{k\rightarrow 0}{%
\lim }\psi _{k}^{\left( 0\right) }\left( x\right) $.

\bigskip We can start to build a chain of isospectral potentials by
successive DBT based on the "fundamental" unphysical eigenfunctions of zero
energy which are recessive at the origin (satisfying the Dirichlet boundary
conditions at the origin). An immediate recurrence gives, for $N_{m}=%
\underset{m\text{ times}}{\underbrace{\left( 0,...,0\right) }}$

\begin{equation}
\left\{ 
\begin{array}{c}
\psi _{0}^{\left( N_{m}\right) }\left( x\right) =x^{m+1} \\ 
V^{\left( N_{m}\right) }\left( x\right) =V_{B}\left( x;m\right)
=m(m+1)/x^{2}.%
\end{array}%
\right.   \label{B1}
\end{equation}

All the extensions $V^{\left( N_{m}\right) }$ subject to the Dirichlet
boundary condition at the origin are then strictly isospectral to $V^{\left(
0\right) }$ with a unique energy band $\left] 0,+\infty \right[ $. These
results are naturally well-known and this constitutes the first example of
application that Darboux gave of his method in his first article on the
subject \cite{darboux2}. They are also one of the key ingredient given by
Duistermaat and Gr\"{u}nbaum \cite{duistermaat} in the proof that every
trivial monodromy potential decaying at infinity can be considered as a
rational extension of the free particle potential.

Nevertheless, the question of the Wronskian representation of the extended
potentials Eq(\ref{B1}) and of their eigenfunctions is less trivial. Indeed,
\ we are now in a case of confluency where all the successive DBT are built
from eigenfunctions associated to the same energy. Then we cannot use the
Crum formulas to express $V^{\left( N_{m}\right) }$ and its eigenstates $%
\psi _{k}^{\left( N_{m}\right) }$ in terms of Wronskians of distinct "seeds
functions" (note that $\widehat{A}(w_{0}^{\left( N_{m}\right) })$ is an
annihilator for $\psi _{0}^{\left( N_{m}\right) }$ and that $\psi
_{0}^{\left( N_{m+1}\right) }$ cannot be considered as the image of $\psi
_{0}^{\left( N_{m}\right) }$ by the DBT $A(w_{0}^{\left( N_{m}\right) })$).
In this case, we can however use Matveev's formulas \cite{matveev,matveev2}
and express $V^{\left( N_{m}\right) }$ and $\psi _{k}^{\left( N_{m}\right) }$
in terms of "generalized Wronskians":

\emph{Matveev's formulas:} In the confluent case where the repeated DBT are
built on eigenfunctions associated to the same value $k_{0}$ of the spectral
parameter $k$ ($N_{m}=\underset{m\text{ times}}{\underbrace{\left(
k_{0},...,k_{0}\right) }\text{)}}$, we can write

\begin{equation}
\left\{ 
\begin{array}{c}
V^{\left( N_{m}\right) }(x)=V\left( x\right) -2\left( \ln W\left( \psi
_{\kappa },\left. \frac{\partial ^{i_{1}}\psi _{\kappa }}{\partial \kappa
^{i_{1}}}\right\vert _{\kappa =k_{0}},...,\left. \frac{\partial
^{i_{m-1}}\psi _{\kappa }}{\partial \kappa {}^{m-1}}\right\vert _{\kappa
=k_{0}}\mid x\right) \right) ^{\prime \prime } \\ 
\\ 
\psi _{k}^{\left( N_{m}\right) }\left( x\right) =\frac{W\left( \psi
_{k_{0}},\left. \frac{\partial ^{i_{1}}\psi _{\kappa }}{\partial \kappa
^{i_{1}}}\right\vert _{\kappa =k_{0}},...,\left. \frac{\partial
^{i_{m-1}}\psi _{\kappa }}{\partial \kappa {}^{m-1}}\right\vert _{\kappa
=k_{0}},\psi _{k}\mid x\right) }{W\left( \psi _{k_{0}},\left. \frac{\partial
^{i_{1}}\psi _{\kappa }}{\partial \kappa ^{i_{1}}}\right\vert _{\kappa
=k_{0}},...,\left. \frac{\partial ^{i_{m-1}}\psi _{\kappa }}{\partial \kappa
{}^{m-1}}\right\vert _{\kappa =k_{0}}\mid x\right) },%
\end{array}%
\right. 
\end{equation}%
where the $\left. \frac{\partial ^{i_{j}}\psi _{\kappa }}{\partial \kappa
^{i_{j}}}\right\vert _{\kappa =k_{0}},\quad j=0,...,m-1,$\ are the $m$ first
non-zero derivative of $\psi _{\kappa }$ with respect to the spectral
parameter $\kappa $ at the value $\kappa =k_{0}$. In our case, for $k_{0}=0$%
, we have

\begin{equation}
\psi _{k}\left( x\right) =\sum_{n=0}^{\infty }\left( -1\right) ^{n}\frac{%
x^{2n+1}}{\left( 2n+1\right) !}\left( k\right) ^{2n},
\end{equation}

The sequence of the $i_{j}$ is given by $i_{j}=2j,\ j\geq 0$ with

\begin{equation}
\left( \frac{\partial ^{2j}\psi _{\kappa }\left( x\right) }{\partial \kappa
^{2j}}\right) _{\kappa =0}=\left( -1\right) ^{j}\frac{x^{2j+1}}{2j+1}
\end{equation}%
and the Matveev formula for the potential gives

\begin{equation}
V(x;m)=2\left( \ln W\left( x,x^{3},...,x^{2m-1}\mid x\right) \right)
^{\prime \prime }.  \label{matpot}
\end{equation}

Darboux's result is readily recovered by using standard properties of the
Wronskians \cite{muir} which allow us to write

\begin{equation}
W\left( x,x^{3},...,x^{2m-1}\mid x\right) =x^{m}\left( \frac{dy}{dx}\right)
^{m\left( m-1\right) /2}\left. W\left( 1,y,...,y^{m-1}\mid y\right)
\right\vert _{y=x^{2}}=x^{m\left( m+1\right) /2}2^{m\left( m-1\right)
/2}\dprod\limits_{j=1}^{m-1}j!,  \label{W11}
\end{equation}%
that is, $V^{\left( N_{m}\right) }(x)=V_{B}(x;m)=m\left( m+1\right) /x^{2}$.

More interesting is the case of the eigenfunctions. Since $\psi _{k}^{\left(
N_{m}\right) }$ is a solution of

\begin{equation}
\psi ^{\prime \prime }\left( x\right) +\left( k^{2}-\frac{m(m+1)}{x^{2}}%
\right) \psi \left( x\right) =0,  \label{Bessel}
\end{equation}%
satisfying the Dirichlet boundary condition at the origin, it can then be
written as

\begin{equation}
\psi _{k}^{\left( N_{m}\right) }\left( x\right) \sim x^{1/2}\mathbf{J}%
_{m+1/2}(kx)=\sqrt{\frac{2}{\pi }}x\mathbf{j}_{m}(kx),  \label{Bessel2}
\end{equation}%
where $\mathbf{J}_{m+1/2}\left( x\right) $ and $\mathbf{j}_{m}\left(
x\right) $ are the usual Bessel and spherical Bessel functions \cite%
{abramowitz,magnus}.

But from the second Matveev's formula we also have

\begin{equation}
\psi _{k}^{\left( N_{m}\right) }\left( x\right) =\frac{W\left(
x,x^{3},...,x^{2m-1},\sin (kx)/k\mid x\right) }{W\left(
x,x^{3},...,x^{2m-1}\mid x\right) }.  \label{epb}
\end{equation}

This can be considered as a Wronskian version of the Rayleigh formula for
Bessel functions \cite{abramowitz}. Indeed, the same handling as before \cite%
{muir} applied to the Wronskian at the numerator of Eq(\ref{epb}) gives

\begin{equation}
W\left( x,x^{3},...,x^{2m-1},\sin (kx)/k\mid x\right) =2^{m\left( m+1\right)
/2}x^{\left( m+1\right) \left( m+2\right) /2}W\left(
1,y,...,y^{m-1},g(y)\mid y\right) ,
\end{equation}%
where $y=x^{2}$ and $g(y)=\sin $c$(kx)=\sin (kx)/kx$. Then

\begin{equation}
W\left( x,x^{3},...,x^{2m-1},\sin (kx)/k\mid x\right) =2^{m\left( m+1\right)
/2}x^{\left( m+1\right) \left( m+2\right) /2}\left(
\dprod\limits_{j=1}^{m-1}j!\right) g^{\left( m\right) }(y)  \label{W2}
\end{equation}%
and with Eq(\ref{W11}), we finally obtain

\begin{equation}
\psi _{k}^{\left( N_{m}\right) }\left( x\right) =\frac{x^{m+1}}{2^{m}}\left( 
\frac{1}{x}\frac{d}{dx}\right) ^{m}\sin \text{c}(kx).
\label{eigenfunctionsm}
\end{equation}

\section{\protect\bigskip Conclusion}

In this letter we have shown how to recover in a simple way the content of
the Gaillard-Matveev theorem which provides a Wronskian representation for
the TDPT potentials. This is directly achieved by combining the Crum formula
and shape invariance arguments applied to specific singular extensions of
the constant potential. The same reasoning can be adapted to obtain
Wronskian representations for the eigenfunctions of the Bessel potentials.
This confluent case necessitates to employ Matveev's generalized Wronskians,
obtaining then a Wronskian version of the Rayleigh formula.

In \cite{gaillard2}, Gaillard and Matveev also consider the case of the
discrete Darboux-P\"{o}schl-Teller potentials (DDPT), for which they give
Casoratian representation formulas. A treatment of these discrete systems by
the preceding approach will be the object of further investigations.

\section{Acknowledgements}

I would like to thank R.\ Milson and D.\ Gomez-Ullate for stimulating
discussions.


\begin{thebibliography}{0}
\expandafter\ifx\csname natexlab\endcsname\relax\def\natexlab#1{#1}\fi
\expandafter\ifx\csname bibnamefont\endcsname\relax
  \def\bibnamefont#1{#1}\fi
\expandafter\ifx\csname bibfnamefont\endcsname\relax
  \def\bibfnamefont#1{#1}\fi
\expandafter\ifx\csname citenamefont\endcsname\relax
  \def\citenamefont#1{#1}\fi
\expandafter\ifx\csname url\endcsname\relax
  \def\url#1{\texttt{#1}}\fi
\expandafter\ifx\csname urlprefix\endcsname\relax\def\urlprefix{URL }\fi
\providecommand{\bibinfo}[2]{#2}
\providecommand{\eprint}[2][]{\url{#2}}

\end{thebibliography}


\begin{thebibliography}{99}
\bibitem{darboux} G.\ Darboux, \textit{Le\c{c}ons sur la th\'{e}orie g\'{e}n%
\'{e}rale des surfaces et les applications g\'{e}om\'{e}triques du calcul
infinit\'{e}simal},\ vol 2, 2nd edition (Gauthier-Villars, Paris, 1915).

\bibitem{gaillard1} P.\ Gaillard and V.\ B.\ Matveev, \textquotedblleft
Wronskian addition formula and its application\textquotedblright , vol. MPI
02-31 (Bonn: Max-Planck-Institut f\"{u}r Mathematik) 1-17 (2002).

\bibitem{gaillard2} P.\ Gaillard and V.\ B.\ Matveev, \textquotedblleft
Wronskian and Casorati determinant representations for Darboux-P\"{o}%
schl-Teller potentials and their difference extensions\textquotedblright ,
J. Phys. A: Math. Gen. \textbf{42}, 404009 (2009).

\bibitem{casahorran} J.\ Casahorran, \textquotedblleft A family of
supersymmetric quantum mechanics models with singular
superpotentials\textquotedblright , Phys. Lett. B \textbf{156}, 425-428
(1991).

\bibitem{panigrahi} P.\ K.\ Panigrahi and U.\ P.\ Sukhatme,
\textquotedblleft Singular superpotentials in supersymmetric quantum
mechanics\textquotedblright , Phys. Lett. A \textbf{178}, 251-257 (1993).

\bibitem{robnik} M.\ Robnik, \textquotedblleft Supersymmetric quantum
mechanics based on higher excited states\textquotedblright ,\ J. Phys. A 
\textbf{30}, 1287-1294 (1997).

\bibitem{marquez} I. F.\ Marquez, J.\ Negro and L.M.\ Nieto,
\textquotedblleft Factorization method and singular
hamiltonians\textquotedblright , J.\ Phys. A: Math. Gen. \textbf{31},
4115-4125 (1998).

\bibitem{marquette} I.\ Marquette, \textquotedblleft Singular isotonic
oscillator, supersymmetry and superintegrability\textquotedblright ,\ SIGMA 
\textbf{8}, 063 (2012).

\bibitem{grandati} Y. Grandati and A. B\'{e}rard, \textquotedblleft Rational
solutions for the Riccati-Schr\"{o}dinger equations associated to
translationally shape invariant potentials\textquotedblright ,\ Ann. Phys. 
\textbf{325}, 1235-1259 (2010).

\bibitem{Ramos} J.\ F.\ Cari\~{n}ena and A.\ Ramos, \textquotedblleft
Integrability of Riccati equation from a group theoretical
viewpoint\textquotedblright ,\ Int. J. Mod. Phys. A \textbf{14}, 1935-1951
(1999).

\bibitem{carinena2} J.\ F.\ Cari\~{n}ena, A.\ Ramos and D.\ J.\ Fernandez,
\textquotedblleft Group theoretical approach to the intertwined
hamiltonians\textquotedblright ,\ Ann. Phys. \textbf{292}, 42-66 (2001).

\bibitem{sukumar2} C.\ V.\ Sukumar, \textquotedblleft Supersymmetry,
factorization of the Schr\"{o}dinger equation and an hamiltonian
hierarchy\textquotedblright , J.\ Phys.\ A \textbf{18}, L57-L61 (1985).

\bibitem{crum} M.\ M.\ Crum, \textquotedblleft Associated Sturm-Liouville
systems\textquotedblright , Q.\ J.\ Math. \textbf{6}, 121-127 (1955).

\bibitem{Matveev} V. B.\ Matveev and M. A.\ Salle, \textit{Darboux
Transformations and Solitons} (Springer-Verlag, Berlin, 1991).

\bibitem{frank} W.\ Frank, D.\ Land and R.\ Spector, \textquotedblleft
Singular potentials\textquotedblright , Rev. Mod. Phys. \textbf{43}, 36-98
(1971). 

\bibitem{lathouwers} L.\ Lathouwers, \textquotedblleft The hamiltonian $%
H=\left( -\frac{1}{2}\right) \frac{d^{2}}{dx^{2}}+\frac{x^{2}}{2}+\frac{%
\lambda }{x^{2}}$ reobserved\textquotedblright , J. Math. Phys. \textbf{16},
1393-1395 (1975). 

\bibitem{cooper} F.\ Cooper, A.\ Khare and U.\ Sukhatme, \textit{%
Supersymmetry in Quantum Mechanics} (World Scientific, Singapore, 2001).

\bibitem{Dutt} R.\ Dutt, A. Khare and U.\ P.\ Sukhatme, \textquotedblleft
Supersymmetry, shape invariance and exactly solvable
potentials\textquotedblright ,\ Am. J. Phys. 5\textbf{6}, 163--168 (1988).

\bibitem{Gendenshtein} L.\ Gendenshtein, \textquotedblleft Derivation of
exact spectra of the Schrodinger equation by means of
supersymmetry\textquotedblright ,\ JETP Lett. \textbf{38}, 356-359 (1983).

\bibitem{darboux2} G.\ Darboux, \textquotedblleft Sur une proposition
relative aux \'{e}quations lin\'{e}aires\textquotedblright , Comptes Rendus
Acad.Sci. \textbf{94}, 1456-1459 (1882).

\bibitem{duistermaat} J.\ J.\ Duistermaat and F.\ A.\ Gr\"{u}nbaum,
\textquotedblleft Differential equations in the spectral
parameter\textquotedblright , Comm. Math. Phys. \textbf{103}, 177-240 (1986).

\bibitem{matveev} V.\ B.\ Matveev, \textquotedblleft Generalized Wronskian
formula for solutions of the KdV equations: first
applications\textquotedblright , Phys. Lett. A \textbf{166}, 205-208 (1992).

\bibitem{matveev2} V.\ B.\ Matveev, \textquotedblleft Positons: slowly
decreasing analogues of solitons\textquotedblright , Theor. Math. Phys. 
\textbf{131}, 483-497 (2002).

\bibitem{abramowitz} M. Abramowitz and I.\ A.\ Stegun, \textit{Hanbook of
mathematical functions} (Dover, New-York, 1972). 

\bibitem{magnus} A. Erd\'{e}lyi, W. Magnus, F. Oberhettinger and F. G.
Tricomi, \textit{Higher transcendental functions (}Mc Graw-Hill\textit{,}
New York, 1953).

\bibitem{muir} T.\ Muir (revised and enlarged by W.H.\ Metzler), \textit{A
treatise on the theory of determinants (}Dover\textit{,} New York, 1960).
\end{thebibliography}
\end{document}